\documentstyle[12pt,epsf]{article}

\begin{document}

\title{On the value of coupling constant.}

\author{
P.I.Pronin
\thanks{E-mail:$pronin@theor.phys.msu.su$}
and K.V.Stepanyantz
\thanks{E-mail:$stepan@theor.phys.msu.su$}}

\maketitle

\begin{center}
{\em Moscow State University, physical faculty,\\
department of theoretical physics.\\
$117234$, Moscow, Russia}
\end{center}

\begin{abstract}
Using an analogy between Bogomol'nyi bound and the harmonic oscillator
in quantum mechanics we propose a possible explanation of the
coupling constant numerical value at the grand unification scale.
It is found to be $1/8\pi$.
\end{abstract}

\sloppy



According to the recent precise experiments \cite{exper}, the running
coupling constants do not meet in a single point in the standard model.
Nevertheless, in its supersymmetric extension (MSSM) \cite{mssm} it
really takes place. (see for example \cite{dienes}). Usually, it was used
as an indirect proof of supersymmetry existence. Nevertheless, this fact can
be also considered from a different point, namely as an indirect measurement
of $\alpha_{GUT}$ (the coupling constant value at the scale of Grand
Unification). Numerically it appeared to be approximately $1/25$.

At the present time there is no theoretical explanation why the coupling
constant has a definite value. In our opinion the Theory Of Everything
should predict this value. Possibly, it is a simple combination of
mathematical constants. Let us note, that numerical value of
$\alpha_{GUT}^{-1}$, produced by the indirect experimental data in MSSM,
is very close to $8\pi \approx 25.13$. Of course, this value depends on
the renormalization scheme beyond the one-loop approximation. Nevertheless,
say, in the Quantum Electrodynamics the measurable values can be obtained
by calculations in the definite scheme. That is why below we will suggest
the possibility of $\alpha_{GUT}$ measurement and the existence of some
preferable (physical) scheme.

In this paper we will also try to seek for a possible explanation
why the coupling constant has a definite value. For this purpose we
will use the analogy between the harmonic oscillator in Quantum Mechanics
and the Bogomol'nyi bound.

Really, the harmonic oscillator is the simplest example of the self-dual
model \cite{harvey}, because its Hamiltonian

\begin{equation}\label{ham}
H = \frac{p^2}{2m} + \frac{m \omega^2 x^2}{2}
\end{equation}

\noindent
is invariant under the transformations

\begin{equation}
p \rightarrow - m \omega x \qquad x \rightarrow p/m\omega
\end{equation}

From the other hand, first duality conjectures \cite{montonen} were
based on the symmetry of Dirac quantization condition and Bogomol'nyi
bound \cite{bg,harvey}

\begin{equation}\label{bogbound}
M \ge v (Q_e^2+Q_m^2)^{1/2}
\end{equation}

\noindent
where $Q_e$ and $Q_m$ are electric and magnetic charges and $M$ is
a mass of state.

Let us note, that the right hand side of (\ref{bogbound}) looks very like
(\ref{ham}).

The commutator of electric and magnetic fluxes through the closed
loops $\alpha$ and $\beta$ respectively is \cite{ashtekar}

\begin{equation}\label{commutator}
[\hat B(\alpha),\hat E(\beta)]=i GL(\alpha,\beta)
\end{equation}

\noindent
where $GL(\alpha,\beta)$ is a Gauss linking number between the loops
$\alpha$ and $\beta$.

Assuming the commutator of electric and magnetic charges to be a limit of
the commutator through the finite contours, we found, that it is not
well defined. In this case we will substitute $GL$ in (\ref{commutator})
by an arbitrary (undefined) integer $m$, so that

\begin{equation}
[\hat Q_e,\hat Q_m]=i m
\end{equation}

The simplest way to satisfy this is to identify

\begin{eqnarray}
&&\hat Q_e = e;\nonumber\\
&&\hat Q_m = \frac{m}{i}\frac{d}{de}
\end{eqnarray}

\noindent
Then, using the analogy with the harmonic oscillator, we write down a
Schrodinger type equation

\begin{equation}\label{eq}
\left(- m^2 \frac{d^2}{de^2} + e^2\right) \psi(e) = 4\pi\alpha \psi(e),
\end{equation}

\noindent
($4\pi$ comes from $\alpha_{class}=e^2/4\pi$ in the chosen unit system)

It is natural to suggest, that $\psi(e)$ is a wave function for the coupling
constant distribution and eigenvalues of the operator in the left hand
are observables (instead of $Q_e^2$ and $Q_m^2$). They are discreet and
equal to $m(1+2n)$. The lowest nontrivial eigenvalue is 1 with the
corresponding eigenfunction

\begin{equation}
\psi_0 = \frac{1}{\pi^{1/4}} exp(-e^2/2).
\end{equation}

Therefore, at the first sight $\alpha=1/4\pi$. Nevertheless, it is
valid only above the Grand Unification Scale. Really, the derivative over
the coupling constant originates from magnetic charge. In Grand Unification
models its mass is about $M_{GUT}$ \cite{monopoles} and, therefore, the
corresponding term is absent in the low energy physics. So, below the
threshold we should omit the operator of squared magnetic charge and
calculate the middle value of

\begin{equation}
\alpha_{GUT} = \frac{<e^2>}{4\pi}
= \frac{1}{4\pi} \int de\ e^2 \frac{1}{\sqrt{\pi}}\ exp(- e^2)
= \frac{1}{8\pi}.
\end{equation}

Then the numerical value of $\alpha^{-1}_{GUT}$ is approximately 25.13.
\footnote{It is easy to see, that the result does not depend on the
particular choice of unit system.},
that agree well with the indirect predictions of MSSM \cite{unf}, if
we take into account recent experimental data \cite{exper}.

As we already mensioned, above the Grand Unification scale

\begin{equation}
\alpha_{GUT} = \frac{1}{4\pi}.
\end{equation}

\noindent
The behavior of running coupling constants is sketched it at the Fig.1.

\noindent
\begin{figure}[h]
\epsfbox{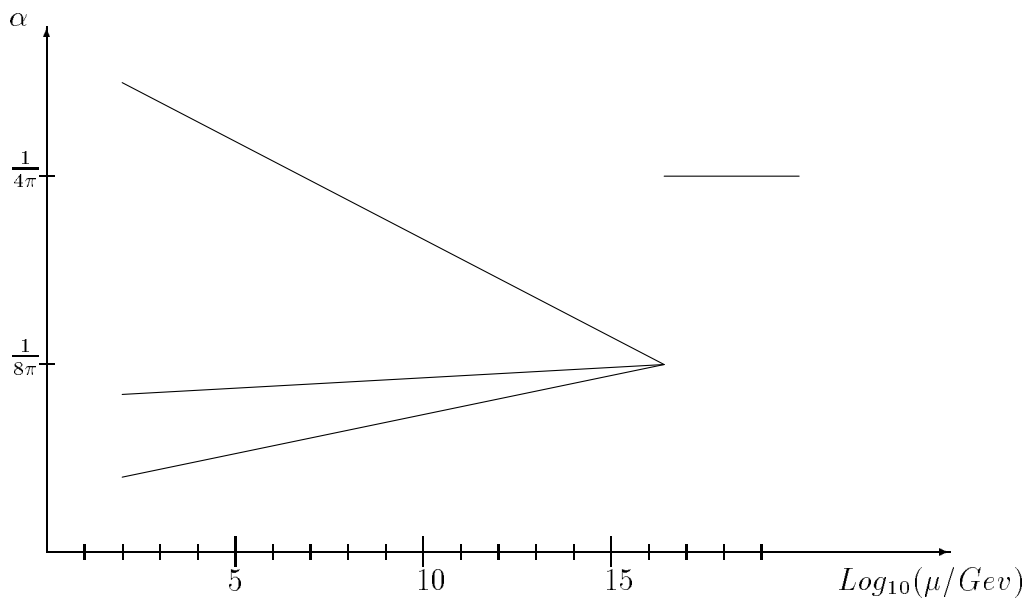}
\caption{Running of the coupling constants.}
\label{fig}
\end{figure}

To conclude we want to mention, that the main ideas of this paper are
rather guesses and analogies, than results. But the hope, that something
may be accidentally true, gave life to the present paper. At least the
equation $\alpha_{GUT}=1/8\pi$ seems rather curious.

\end{document}